\documentclass{aa}  

\usepackage{graphicx}
\usepackage{txfonts}
\usepackage{xcolor}
\usepackage{amsmath,amssymb}
\usepackage{textcomp}
\newcommand{\VIK}{VIK~J2318$-$3113}
\newcommand{\VIKs}{VIK~J2318$-$3113 }
\newcommand{\mbeam}{mJy~beam$^{-1}$}

\begin{document}

   \title{Constraining the Radio Properties of the $z$=6.44 QSO VIK~J2318$-$3113}

   \author{Luca Ighina
          \inst{1,2}
          \and
          James K. Leung\inst{3,4,5} \and Jess W. Broderick\inst{6} \and Guillaume Drouart\inst{6} \and Nick Seymour \inst{6}
          \and
          Silvia Belladitta\inst{1,2} \and Alessandro Caccianiga\inst{1} \and Emil Lenc\inst{4}
          \and Alberto Moretti\inst{1}
          \and Tao An\inst{8,9} \and Tim J. Galvin\inst{6,10}  \and George H. Heald\inst{10} \and Minh T. Huynh\inst{10}  \and David McConnell\inst{4} \and Tara Murphy\inst{3,5} \and Joshua Pritchard\inst{3,4,5} \and Benjamin Quici\inst{6} \and Stas S. Shabala\inst{11,12} \and Steven J. Tingay\inst{6} \and Ross J. Turner\inst{11} \and Yuanming Wang\inst{3,4,5} \and Sarah V. White\inst{13}
            }

   \institute{
    INAF, Osservatorio Astronomico di Brera, via Brera 28, 20121, Milano, Italy\\
    \email{lighina@uninsubria.it}
    \and
    DiSAT, Universit\`a degli Studi dell'Insubria, via Valleggio 11, 22100 Como, Italy
    \and
    Sydney Institute for Astronomy, School of Physics, University of Sydney, NSW 2006, Australia
    \and
    CSIRO Space and Astronomy, PO Box 76, Epping, NSW, 1710, Australia
    \and 
    ARC Centre of Excellence for Gravitational Wave Discovery (OzGrav), Hawthorn, VIC 3122, Australia \and
    International Centre for Radio Astronomy Research, Curtin University, 1 Turner Avenue, Bentley, WA, 6102, Australia
    \and
    ARC Centre of Excellence for All-sky Astrophysics (CAASTRO), The University of Sydney, NSW 2006, Australia
    \and
    Shanghai Astronomical Observatory, Chinese Academy of Sciences, Nandan Road 80, Shanghai 200030, China
    \and
    Key Laboratory of Cognitive Radio and Information Processing, Guilin University of Electronic Technology, 541004 Guilin, China
    \and
    CSIRO Space and Astronomy, PO Box 1130, Bentley WA 6102, Australia
    \and
    School of Natural Sciences, University of Tasmania, Private Bag 37, Hobart, TAS 7001, Australia
    \and
    ARC Centre of Excellence for All-Sky Astrophysics in 3 Dimensions (ASTRO 3D)
    \and
    Department of Physics and Electronics, Rhodes University, PO Box 94, Makhanda, 6140, South Africa
    }

   \date{Received 24 November 2021; accepted 11 March 2022}

  \abstract{
   The recent detection of the quasi-stellar object (QSO) VIKING J231818.3$-$311346 (hereafter \VIK) at redshift $z=6.44$ in the Rapid ASKAP Continuum Survey (RACS) uncovered its radio-loud nature, making it one of the most distant known to date in  this class. By using data from several radio surveys of the Galaxy and Mass Assembly 23$^\mathrm{h}$ field and from dedicated follow-up, we were able to constrain the radio spectrum of \VIKs in the observed range $\sim$0.1--10~GHz. At high frequencies (0.888--5.5~GHz in the observed frame) the QSO presents a steep spectrum ($\alpha_{\rm r}$=1.24, with $S_\nu\propto \nu^{-\alpha_{\rm r}}$), while at lower frequencies (0.4--0.888~GHz in the observed frame) it is nearly flat. The overall spectrum can be modelled by either a curved function with a rest-frame turnover around 5~GHz, or with a smoothly varying double power law that is flat below a rest-frame break frequency of about 20~GHz and which significantly steepens above it. Based on the model adopted, we estimated that the radio jets of \VIKs must be a few hundred years old, in the case of a turnover, or less than few$\times$10$^4$ years, in the case of a break in the spectrum.
   Having multiple observations at two frequencies (888~MHz and 5.5~GHz), we further investigated the radio variability previously reported for this source. We found that the marginally significant flux density variations are consistent with the expectations from refractive interstellar scintillation, even though relativistic effects related to the orientation of the source may still play a non-negligible role. Further radio and X-ray observations are required to conclusively discern the nature of this variation.}


   \maketitle
%

\section{Introduction}

High-redshift quasi-stellar objects (QSOs) are among the most important tools to investigate the properties of supermassive black holes (SMBHs) in the initial stages after their formation. QSOs able to expel part of the accreting material in the form of two bipolar relativistic jets (radio loud, RL\footnote{From the observational point of view we define a source to be radio loud if the rest-frame flux density ratio $S_\mathrm{5GHz}$/$S_\mathrm{4400\AA} \textgreater10$; \cite{Kellerman1989}. Even though this threshold has been shown to be rather simplistic in the separation between jetted and non-jetted QSOs \citep[e.g.][]{Padovani2017}, throughout the paper we use the terms jetted and RL interchangeably for simplicity and consistency with other works.}) are of particular interest since the presence of such jets can significantly affect the growth of the SMBH itself and also its environment \citep{Blandford2019}. While optical and infra-red (IR) observations can be used to investigate the accretion process \citep[e.g.][]{Tang2019,Eilers2021} and its re-ionising effect on the surrounding intergalactic medium \citep[e.g.][]{Becker2019}, radio and X-ray data are crucial to study the properties of the relativistic jets \citep[e.g.][]{Ighina2022} and their specific effect on the growth of SMBHs, the host galaxies and the environments \citep[e.g.][]{Croston2019, Rojas-Ruiz2021}.

As a result of ongoing and upcoming wide-area radio surveys with the Square Kilometre Array (SKA; \citealt{Braun2015}) and its precursors (e.g. \citealt{Norris2021,Murphy2021}), we will be able to use deep multi-band radio data to discover and study statistical samples of RL sources even at the highest redshifts ($z$\textgreater6), where only a handful of RL Active Galactic Nuclei (AGNs) have been found so far (see e.g. Tab. 6 in \citealt{Banados2021}). The systematic characterisation of these objects, such as their spectral shape and their short/long-term variability, will provide unique information on the first phases of the launch and propagation of relativistic jets from young SMBHs (e.g. \citealt{Nyland2020}) and their connection with the powerful large-scale radio galaxies observed in the local Universe (e.g. \citealt{Alexander2000,Oei2022}).

The first $z$\textgreater6 QSO whose RL nature was uncovered because of the new generation of all-sky radio surveys is VIKING J231818.3$-$311346 (hereafter \VIK; \citealt{Ighina2021}), discovered in the Rapid ASKAP Continuum Survey (RACS; \citealt{McConnell2020}). In this work we present the radio properties of this source, focusing on its spectral shape and variability. The work is based on radio data obtained with the Murchison Widefield Array (MWA; \citealt{Tingay2013}), the Australia Telescope Compact Array (ATCA; \citealt{Wilson2011}) and the Australian Square Kilometre Array Pathfinder \citep[ASKAP;][]{Johnston2007,Hotan2021} as part of the Galaxy And Mass Assembly (GAMA; \citealt{Driver2011}) 23$^{\rm h}$ field and through dedicated follow-up observations.

Throughout the paper we assume a flat $\Lambda$CDM cosmology with $H_{0}$=70 km s$^{-1}$ Mpc$^{-1}$, $\Omega_\mathrm{M}$=0.3 and $\Omega_{\Lambda}$=0.7, where 1\arcsec\, corresponds to a projected distance of 5.49~kpc at $z$=6.44. Spectral indices are given assuming $S_{\nu}\propto \nu^{-\alpha}$ and all errors are reported at 68\% confidence unless specified otherwise.

\section{Multiwavelength Spectral Energy Distribution of \VIK}

Thanks to the multifrequency data available for \VIK, we were able to constrain its spectral energy distribution (SED) from the radio ($\sim$10$^{9}$~Hz) to the optical band ($\sim$10$^{15}$~Hz). The SED is shown in Fig. \ref{fig:rad_spec}. Optical/near-IR data ($Z$, $Y$, $J$, $H$ and $K_\mathrm{s}$ bands; 0.1--0.3~\textmu m in the source's rest frame) from the VISTA Kilo-Degree Infrared Galaxy Survey (VIKING; \citealt{Edge2013}) and the $W1$--$W2$ IR bands (0.5--0.6~\textmu m rest frame) from the \textit{Wide-field Infrared Survey Explorer} \citep{Wright2010} catalogue (catWISE; \citealt{Eisenhardt2020}) are reported as orange and brown triangles. Besides the detections in these surveys, \VIKs has also been targeted with the Atacama Large Millimetre/submillimetre Array (ALMA; \citealt{Wootten2009}). Specifically, the source was first observed in band 6 (247~GHz, $S_\nu$=0.36$\pm$0.08~mJy; see \citealt{Decarli2018} and \citealt{Venemans2020}) and then in band 3 (102~GHz, $S_\nu$=0.05$\pm$0.01~mJy; P.I. R. Decarli, project code: 2019.1.00147.S). This last value is the peak flux density measured in the calibrated image obtained from the standard reduction ALMA pipeline. We report these data, which correspond to the 750--1900~GHz or 160--400~\textmu m range in the source rest frame, as blue squares in Fig.\ref{fig:rad_spec}. 

We normalised the sum of the QSO and the starburst galaxy (M82) templates from \cite{Polletta2007} to the available optical/IR observations and then extended to lower radio frequencies assuming a power law with a spectral index $\alpha_\mathrm{r}^\mathrm{SF}$=0.75 \citep{Condon2002}. The corresponding shaded area represents a different normalisation of the template that takes into account the uncertainty on the ALMA data and also a different slope of the radio emission ($\alpha_{\rm r}^{\rm SF}$=0.7--0.8; \citealt{Perez2021}). Based on the expected contribution from the synchrotron emission in star-forming (SF) galaxies (see Fig. \ref{fig:rad_spec}), it is clear that the observed radio emission in \VIKs is dominated by the AGN activity. Therefore, the observed features of the radio spectrum should be directly linked to the radiation produced within the jet and its characterisation can provide useful insight into its properties.

\section{Radio data}
The detection of \VIKs at 888~MHz in the first data release of the low-band RACS survey revealed its RL nature. Nevertheless, this source is too faint to be detected in the majority of publicly available radio surveys. As reported in \cite{Ighina2021}, \VIKs is not detected in the Tata Institute of Fundamental Research Giant Metrewave Telescope (GMRT) Sky Survey (TGSS; \citealt{Intema2017}), the Sydney University Molonglo Sky Survey (SUMSS; \citealt{Bock1999,Mauch2003,Murphy2007}), and the National Radio Astronomy Observatory Very Large Array Sky Survey (NVSS; \citealt{Condon1998}). The only detection is at 3~GHz (signal-to-noise ratio, S/N=3) from the second epoch of the Karl G. Jansky Very Large Array (JVLA) Sky Survey (VLASS; \citealt{Lacy2020}).

After the publication of \cite{Ighina2021}, the official catalogue of the first low-band epoch of the RACS survey was published \citep{Hale2021}. This catalogue was built by combining all the tiles between $-$80$^\mathrm{o}$\textless Dec\textless30$^\mathrm{o}$ and by convolving them to a common resolution of 25\arcsec. In this work we adopt the values of \VIKs reported in this catalogue as best-fit measurements of the RACS low-band observations ($S^\mathrm{peak}_\mathrm{888MHz}$=1.17$\pm$0.23 \mbeam \, and RMS = 0.22~\mbeam). These values are slightly different, but still consistent, compared to the ones reported in \cite{Ighina2021}. This is because the value reported in \cite{Hale2021} is the combination of multiple observations where \VIKs is on the edge of the field of view.

In this section we give a brief description of further observations available for \VIK, either as part of the GAMA project or through dedicated follow-up observations. A summary of the data used in this work is given in Tab. \ref{Tab:Observations} and we present the corresponding radio images in Appendix \ref{sec:radio_images}. We also include a new observation from the RACS mid-band survey (at 1.37~GHz, as opposed to low-band at 888~MHz). The overall data reduction for the second release of the RACS survey is still in progress, but it is very similar to the process carried out for the first low-band data release (see \citealt{McConnell2020} for a description). 

\begin{table*}
\centering
\caption{Summary of all the observations available for \VIK. The images of previously unpublished radio observations are reported in appendix \ref{sec:radio_images}. The source is not resolved in any image, the position angles are measured from North to East and all the upper limits are at the 3$\sigma$ level.}
\label{Tab:Observations}

\begin{tabular}{lcccccccc}
    \hline
    \hline
    Project/Survey & Telescope & Frequency & Bandwidth & RMS & Beam sizes & P.A. & Obs. Date & Peak flux density\\
    & & MHz & MHz & \textmu Jy~beam$^{-1}$ & maj\arcsec$\times$min\arcsec & deg & y-m-d &\mbeam\\
    \hline
    G23--MIDAS  & MWA   & 154   & 30.72     & 990   & 95$\times$70      & 154.5 & 2017-11-02 & \textless3.3\\
                & MWA   & 216   & 30.72     & 540   & 67$\times$51      & 154.0 & 2017-11-02 & \textless 1.8\\
    G23--GLASS  & uGMRT & 399   & 200       & 105   & 12.4$\times$8.0   & 15.0  & 2017-03-03 & 0.89 $\pm$ 0.15\\
                & ATCA  & 5500  & 2048      & 32    & 5.5$\times$3.5    & 0     & 2018-02-20 & 0.104 $\pm$ 0.034\\
    G23--EMU    & ASKAP & 888   & 288       & 39   & 10.2$\times$8.5    & 82.6  & 2019-03-08 & 0.59 $\pm$ 0.09\\
           
    \\
    Dedicated  & ASKAP & 888 & 288 & 33 & 13.6$\times$10.8 & 105.5 & 2021-02-28 & 1.00 $\pm$ 0.11\\
    Simultaneous & ATCA & 2100 & 2048 & 36 & 17.5$\times$3.9 & 31.0 & 2021-02-28 & 0.46 $\pm$ 0.09 \\
    Follow-up & ATCA & 5500 & 2048 & 20 & 7.6$\times$1.4 & 25.7 & 2021-02-28 & 0.100 $\pm$ 0.022\\
    Observations & ATCA & 9000 & 2048 & 17 & 5.9$\times$1.1 & 27.5 & 2021-02-28 & \textless0.056\\
    
    \\
   Public Surveys:\\
    TGSS & GMRT & 148 & 17 & 2970 & 39.8$\times$25.0& 0 & 2016-03-15 &\textless 8.9 \\
    SUMSS & MOST & 843 & 3 & 2500 & 84.9$\times$45.0& 0 & 2006-09-11 &\textless 7.5\\
    RACS-low    & ASKAP & 888 & 288 & 220 & 25$\times$25 & 0 & 2020-03-27 & 1.17 $\pm$ 0.23\\
    RACS-mid    & ASKAP & 1367 & 144 & 160 & 9.2$\times$7.6 & 69.1 & 2021-01-06 & 0.52 $\pm$ 0.17\\
    NVSS & JVLA & 1435 & 42 & 453 & 45$\times$45 & 0 & 1993-10-09 & \textless 1.36\\
    VLASS\_1 & JVLA & 3000 & 2000 & 114 & 2.7$\times$1.9 & 135.8 & 2018-02-13 & \textless 0.39\\
    VLASS\_2 & JVLA & 3000 & 2000 & 132 & 2.7$\times$2.0 & 10.0 & 2020-11-01 & 0.40  $\pm$ 0.15\\

    \hline
    \hline
\end{tabular}
\end{table*}

\subsection{GAMA 23 field radio observations}

\VIKs is located in the GAMA 23$^{\rm h}$ field (G23), which has been covered by several deep radio observations spanning a large range of frequencies, from $\sim$0.1 to $\sim$5~GHz. 
The G23 field was observed with ASKAP as part of the Evolutionary Map of the Universe \citep[EMU;][]{Norris2011} Early Science Project; details on these observations relevant to \VIKs are reported in \cite{Ighina2021}.

The lower-frequency observations (154--216~MHz) have been performed with the MWA in its extended configuration (MWA Phase II; \citealt{Wayth2018}) as part of the MWA Interestingly Deep Astrophysical Survey (MIDAS; Quici et al. in prep.). Depending on the frequency, the RMS reached by the observations is 0.99~\mbeam \, at 154~MHz and 0.54~\mbeam \, at 216~MHz, while the spatial resolution is between 1\arcmin \, and 1.5\arcmin \, for both images. In the following, we adopt the conservative value of 8\% for the uncertainty on the flux scale, which is based on the comparison between the MIDAS images and the GaLactic and Extra-galactic All-sky MWA (GLEAM; \citealt{Wayth2015}) survey (extra-galactic data release: \citealt{Hurley-walker2017}). See sec. 2.1.2 in \cite{Quici2021} for further details.

As part of the GAMA Legacy ATCA Sky Survey (GLASS), the G23 field has also been covered at 399~MHz with the upgraded GMRT (uGMRT; \citealt{Gupta2017}), and at 5.5~GHz with the ATCA. The 399-MHz image reaches a sensitivity of 0.1~\mbeam \, and a resolution of $\sim$10\arcsec \, near \VIK.  In addition, the ATCA observations were conducted with the extended 6A configuration, reaching a resolution of $\sim$4\arcsec\,  and an RMS of 32~\textmu Jy~beam$^{-1}$ at 5.5~GHz.
We adopt an uncertainty on the absolute flux scale of 10\% and 5\% for the uGMRT and ATCA images respectively.

\subsection{Dedicated Simultaneous Follow-up Observations}

Given the relatively large and rapid variability observed at 888~MHz in \cite{Ighina2021}, we also carried out simultaneous follow-up at different frequencies with the ASKAP and ATCA telescopes on 2021 February 28. The ASKAP observation was reduced using the \texttt{ASKAPsoft} data-reduction pipeline \citep{Guzman2019} and it is publicly available in the CSIRO ASKAP Science Data Archive\footnote{https://data.csiro.au/collections/domain/casdaObservation/search/} (CASDA; \citealt{Huynh2020}) under the identification code SBID 23721. Similarly to the first data release of the RACS survey, it has a bandwidth of 288~MHz centred at 888~MHz. The angular resolution is $\sim$12\arcsec \, and the off-source RMS is 33~\textmu Jy~beam$^{-1}$. We adopt an uncertainty of 10\% on the absolute flux scale.

The follow-up with the ATCA was carried out in the 6D array configuration at three different frequencies: 2.1, 5.5 and 9~GHz (project CX481). The corresponding restored beam sizes (RMS) are respectively: 17.5\arcsec$\times$3.9\arcsec (36~\textmu Jy~beam$^{-1}$), 7.6\arcsec$\times$1.4\arcsec (20~\textmu Jy~beam$^{-1}$) and 5.9\arcsec$\times$1.1\arcsec (17~\textmu Jy~beam$^{-1}$). The absolute flux scale calibration was made with the PKS~1934$-$638 and the resulting uncertainty that we adopt is 5\%.

   \begin{figure*}
   \centering
   \includegraphics[width=0.53\linewidth]{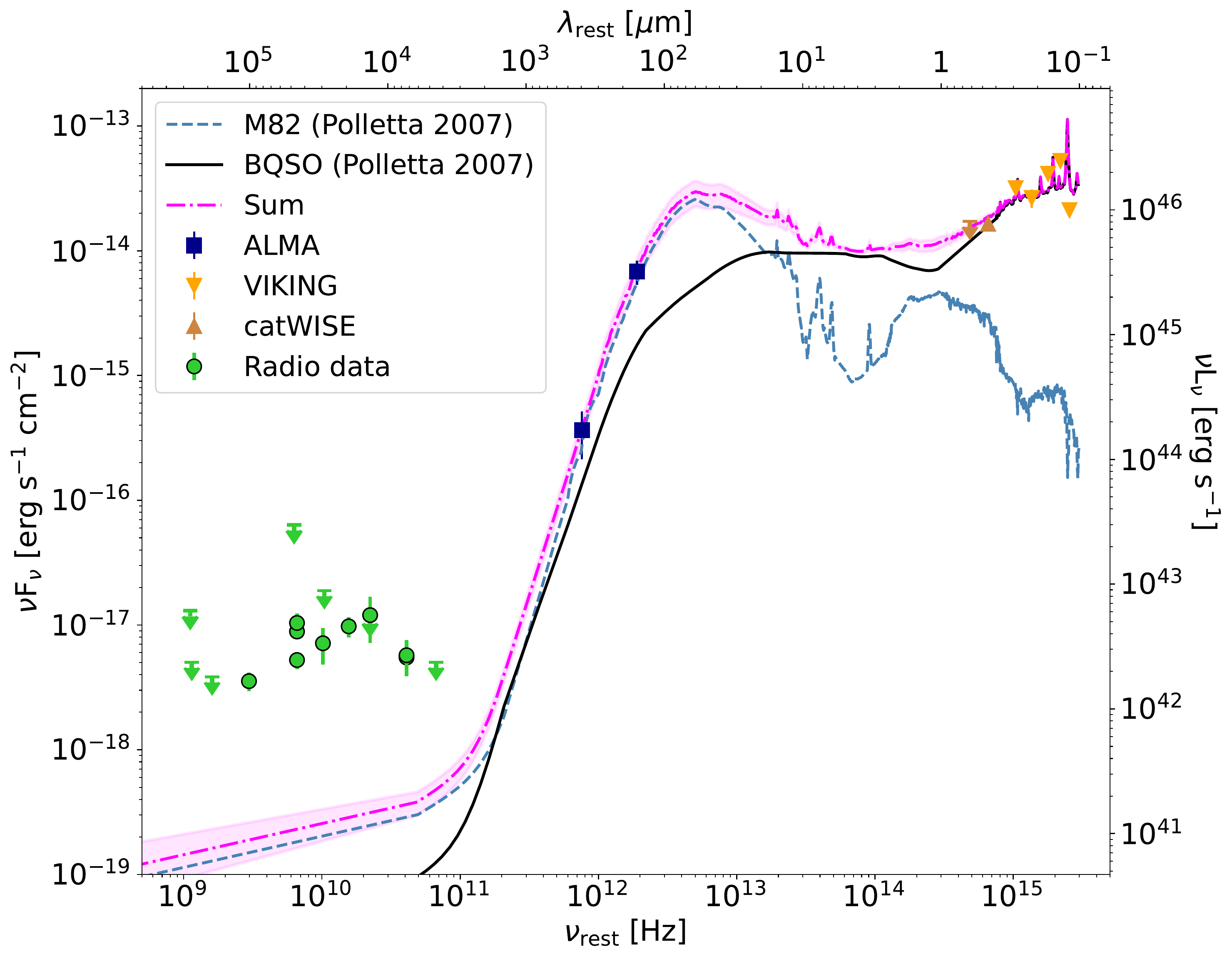}
    \includegraphics[width=0.46\linewidth]{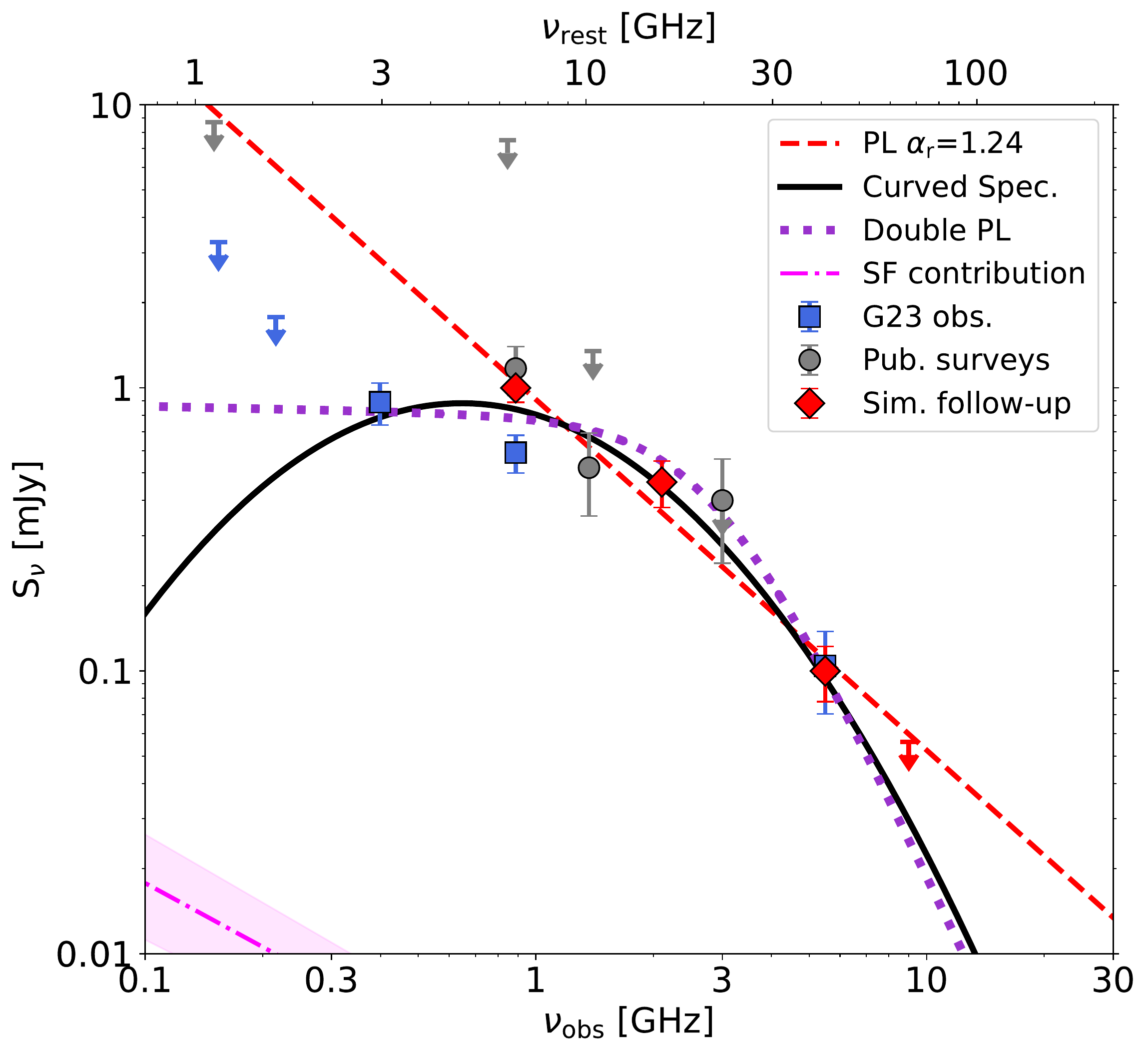}

   \caption{{\bf Left panel:} radio-to-optical/UV rest-frame SED of \VIK. The solid black line and the dashed blue line are two templates of a QSO and a starbust galaxy from \cite{Polletta2007}. Their sum (dashed--dotted magenta line) is normalised to the VIKING--catWISE (orange and brown triangles) and the ALMA (blue squares) data points and then extended in the radio band with a $\alpha_\mathrm{r}^\mathrm{SF}$=0.75 power-law. The radio data are shown as green points and arrows (when upper limits).
   {\bf Right panel:} Radio spectrum of \VIK, where the different data points are divided according to the project under which they were obtained: G23 (blue squares), simultaneous follow-up (red diamonds) and public surveys (grey circles). 3$\sigma$ upper limits are shown as downwards arrows. The dashed red line represents the single power law that best describes the simultaneous detections, $\alpha_\mathrm{r}$=1.24, while the solid black (dotted purple) line is the curved spectrum (double power law) obtained from the fit of all the radio data available with the \texttt{MrMoorse} code \citep{Drouart2018}. The magenta dashed-dotted line is the expected contribution from SF to the radio emission, as reported in the left panel.}
    \label{fig:rad_spec}
    \end{figure*}

\section{Radio spectrum}
\label{sec:radio_spec}

In Fig. \ref{fig:rad_spec} (right panel) we show all the radio measurements available for \VIKs based on the projects under which the single observations were taken.
The VLASS data at 3~GHz come from the quick-look images of the survey, which have been shown to be potentially unreliable at faint fluxes. For this reason we add a further 15\% uncertainty on these measurements, see \cite{Gordon2021} for further details. 
Upper limits are at the 3$\sigma$ level, and \cite{Baars1977} is our reference flux density scale. Where needed, we account for the offset between the \cite{Baars1977} scale and other flux density scales by scaling the total flux density and including an appropriate uncertainty in the estimate of their error.
Based on the simultaneous detections only, the radio spectrum of \VIKs can be described by a single power law with a slope $\alpha_{\rm r}$=1.24$\pm$0.04 (dashed red line in Fig. \ref{fig:rad_spec}) between the 888~MHz and 5.5~GHz observed frequencies (corresponding to the rest-frame range $\sim$7--40~GHz). This value is consistent to what is normally found in other RL QSOs at similar frequencies (e.g. \citealt{Bruni2012}), even when considering the few at $z$\textgreater6 \citep{Frey2011, Banados2021}. However, the non-detection at 9~GHz might imply a steepening of the spectrum at higher frequencies, again, in a similar way to what has been observed in other high-redshift radio sources \citep[e.g.][]{Drouart2020,An2022}.

Even though the best way to characterise the shape of the intrinsic spectrum would be from simultaneous data points, in order to constrain a much larger range of frequencies, we also consider non-simultaneous observations with the caveat that the following considerations may vary based on the degree of variability of the source (see sec. \ref{sec:variability}).
While the available data between 0.888 and 5.5~GHz (in the observed frame) are consistent with the steep spectrum inferred from the simultaneous measurements, the non-detections from the MWA and TGSS surveys, together with the measurement from the GLASS--uGMRT seem to indicate a flattening of the radio spectrum at lower frequencies. This feature could be related either to the presence of a turnover at low frequencies or to a flat radio spectrum that steepens at higher frequencies. In the following we discuss both possibilities together with their physical implications.

\subsection{Curved spectrum}
A turnover in the radio spectrum has already been observed in many objects, referred to as ``peaked-spectrum'' (PS) radio sources, and is normally associated with the young age of the radio jet responsible for the radio emission \citep[e.g.][]{Odea2021}.
The most likely explanations for the presence of the turnover are synchrotron self-absorption (SSA) or free--free absorption (FFA). While determining the specific mechanism that is responsible for the turnover in a given source can be of great importance in understanding either its intrinsic properties or the external environment in which the source resides, it is often hard to uniquely distinguish between the two scenarios especially as they could both occur at the same time. Therefore, in order to constrain the frequency at which the potential turnover occurs in \VIK, we performed a fit to the overall radio data available with a generic curved spectrum, given by:
\begin{equation}
\centering
    S_\nu = N \, \nu^{-\alpha} \, e^{q\,({\mathrm{ln}} \nu)^2},
    \label{eq:curved_pl}
\end{equation}
where $N$ is the normalisation and the $q$ parameter is a measurement of the curvature of the spectrum ($|q|$\textgreater0.2 characterises a significantly curved spectrum; \citealt{Callingham2017}) and is related to the peak frequency as $\nu_\mathrm{peak}$ = $e^{\alpha/2q}$. 

We performed the fit in the source's rest frame using the \texttt{MrMoose}\footnote{https://github.com/gdrouart/MrMoose} code (see full description in \citealt{Drouart2018,Drouart2018a}), which takes into account the upper limits available for \VIKs (especially from the MWA and ATCA at \textless200~MHz and 9~GHz, respectively).
The best-fit function is reported as a solid black line in Fig. \ref{fig:rad_spec}.
With this model the resulting peak frequency of \VIKs is at 4.8$\pm$0.6~GHz ($q$=$-$0.49$^{+0.06}_{-0.07}$ and $\alpha$=$-$1.54$^{+0.29}_{-0.34}$), which corresponds to about 650~MHz in the observed frame, similar to the values observed in other $z$\textgreater5 RL QSOs \citep[e.g.][]{Shao2021}.

Several studies have now consolidated the relation between the peak frequency and linear size in these PS systems (e.g. \citealt{Jeyakumar2016}), where the higher the peak frequency the more compact (i.e. younger) the radio jet is. Therefore, we can use the turnover frequency derived from the curved spectrum fit above and the relation derived in \cite{Orienti2014} to obtain a rough estimate (given the large scatter of the relation; e.g. \citealt{Nyland2020}) of the expected linear size of the relativistic jets in \VIK: $\sim$30~pc . 
If we consider an average expansion velocity of $\sim$0.2$c$ \citep[typical for jets with similar luminosities, e.g.][]{An2012}, the corresponding kinetic age of the radio jet would be $\sim$500~years, that is, a relatively newborn jet \citep[e.g.][]{Gugliucci2005}.

\subsection{Double power law}
Besides a curved spectrum, which could be physically motivated for young radio sources, we also used a smooth double power law model to fit the radio emission of \VIK. In this case the steepening of the spectrum at higher frequency could be related to the radiative cooling  of the most energetic electrons (e.g. through the synchrotron or inverse Compton, IC, processes). The double power law model we used is given by:
\begin{equation}
    S_\nu = N \, \nu^{-\alpha_{\rm low}} \, \left[\,  1 + \left(\frac{\nu}{\nu_{\rm b}}\right)^{|\alpha_{\rm low}-\alpha_{\rm high}| } \right]^{\,{\rm -sgn}( \alpha_{\rm high}-\alpha_{\rm low})},
\end{equation}
in this case we found that the best-fit spectrum (again using the \texttt{MrMoose} code) would be nearly flat ($\alpha_{\rm low}$=0.03$^{+0.19}_{-0.27}$) below the break frequency ($\nu_{\rm b}$=20.5$\pm7.1$~GHz, in the rest frame) and then it would sharply steepen afterwards ($\alpha_{\rm high}$=2.92$_{-0.74}^{+1.06}$). The best-fit double power law is reported in Fig. \ref{fig:rad_spec} as a dotted purple line. It is worth highlighting the significant steepening of the radio spectrum that is needed in this scenario, $\alpha_{\rm high} - \alpha_{\rm low}\sim3$. Given the very high redshift of \VIK, this could be the result of the IC scattering off the cosmic microwave background (CMB) \citep[e.g.][]{Ghisellini2015}. Indeed, at $z$=6.44  the equivalent magnetic field associated with the CMB energy density is $B_{\rm IC}$=176~\textmu G, which could potentially dominate the energy losses of high-energy electrons.

From the value of the potential break frequency and a reasonable assumption of the magnetic field strength, we can infer the radiative age of the radio jet and its electron population, see e.g. eq. 1 in \cite{Morabito2016}.
Given the rather strong dependence of the radiative age on the assumed magnetic field, we can consider two main scenarios.
If the radio emission of \VIKs is dominated by the most extended regions, where the magnetic field is weaker, the observed spectrum will be dominated by IC/CMB losses. In this case, assuming a typical magnetic field of the order of $\sim$10~\textmu G \citep[e.g.][]{Croston2005} the corresponding radiative age is $\sim$4$\times$10$^4$~yr. If instead the observed radio emission is produced in very compact regions (e.g. in small hotspots or in the vicinity of the SMBH) the magnetic field strength can even be \textgreater1~mG \citep[e.g.][]{Keim2019}, which would correspond to a radiative age \textless10$^4$~yr. 
Finally, the maximum value of the radio jets age, $\sim$9$\times$10$^4$~yr, is obtained for moderate magnetic fields $\sim$100~\textmu G (1/$\sqrt{3}$ times the $B_{\rm IC}$) and therefore for physical conditions in between the two previous scenarios. Based on the range of radiative age values just derived, \textless10$^5$~yr, we can conclude that \VIKs is likely a relatively young radio source similar to most compact steep spectrum radio sources \citep[e.g.][]{Murgia1999}.\\

Since both the curved spectrum and double power law can well fit our current data and can be physically motivated, we cannot determine which spectral shape is more representative of the intrinsic emission of \VIK. Simultaneous follow up observations covering a broad frequency range (e.g. 0.1--10~GHz) will be needed to constrain its spectrum and thus understand the physical mechanisms responsible for it.



\section{Variability}
\label{sec:variability}

In \cite{Ighina2021}, by comparing the ASKAP images of \VIKs at 888~MHz from RACS and the G23 project, we found that the radio emission of the source might be highly variable. Having an additional deep observation at 888~MHz and two new data points at 5.5~GHz, we can now put further constraints on the variability of \VIK. In particular, from Fig. \ref{fig:rad_spec} and Tab. \ref{Tab:Observations} it seems that variability is especially relevant at lower frequencies, where the flux density estimates at 888~MHz are inconsistent. In contrast, at 5.5~GHz the data points (even if only two) are fully consistent. 
For the moment, we can only investigate the variability of the source in the radio band, since there are no multi-epoch observations in other electromagnetic bands. Further multiwavelength observations will allow us to expand this study and to have a more comprehensive understanding of this source.


In order to see if variability was still relevant with the new available datasets, we checked the consistency of the G23 image with the new RACS catalogue from \cite{Hale2021} as well as with the one available from the dedicated follow-up, which has sensitivity similar to the G23 observation (RMS$\sim$0.04~mJy~beam$^{-1}$ ). Following the previous approach \citep{Ighina2021}, we compared the flux densities of nearby sources ($\lesssim$1 deg from \VIK) above 1~mJy (when considering the RACS catalogue) and above 0.2~mJy (when comparing the G23 and the follow-up images) for sources unresolved both in the G23 and follow-up images ($S_{\rm int}$/$S_{\rm peak}$ \textless 1.3).
We note that while for the G23 and follow-up images we considered the integrated flux densities obtained from a 2D Gaussian fit performed with the Common Astronomy Software Applications \citep[CASA;][]{Mcmullin2007}, for the estimates from the RACS catalogue we considered the peak flux densities, since at faint fluxes the 2D gaussian fit may not be reliable (see section 5 and Fig. 8 in \citealt{Hale2021}). From the comparison of the three images, we found that the distributions of the flux density ratios indicate that the different calibrations are consistent with
each others with offsets always \textless$\pm$0.1~mJy (i.e. smaller than the statistical errors on the estimates).

We note that the observed discrepancy between the G23 flux density and the other measurements is at about 2.5--3$\sigma$ level and therefore further observations are needed to ultimately confirm the physical nature of the variability in the radio emission of \VIK. However, if we interpret this discrepancy as true variability of the source and not a statistical fluctuations, current data suggest that it is not due to calibration effects, but rather it must be either an intrinsic variation (i.e. related to a change in the jet's properties) or extrinsic (i.e. related to a phenomenon external to the AGN and its jet), or a combination of both. In the following, both scenarios will be discussed and evaluated.

In order to quantify the amount of variability present in \VIKs, we use the modulation index, defined as  $M$=$\sigma$/$\bar{S}$, where $\sigma$ is the standard deviation and $\bar{S}$ is the mean of the observed flux densities. At 888~MHz, we find that $M$=33\% ($\sigma$=0.30~\mbeam and $\bar{S}$=0.92~\mbeam); while if we consider the debiased modulation index (e.g. eq. 6. in \citealt{Bell2014}) we obtain $M_\mathrm{deb}$=22\%.

\subsection{Intrinsic Variability}

If the flux variation at 888~MHz is intrinsic to the jet, we can obtain a qualitative estimate of the maximum size of the jet's variable component by using the light-crossing time: $r = \tau c$. The time elapsed between the minimum and maximum flux densities of \VIKs is $\tau$ = 50~days (about one year in the observed frame), which corresponds to a size of \textit{r} $\leq$ 0.04~pc. This value is only a rough upper-limit of the actual size of the varying component in the jet, since it is based on the observation dates, which may not be representative of the intrinsic timescales of the source.

Assuming that the SMBH hosted in \VIKs has a mass of $\sim$10$^9$~M$_\odot$ (as is typically observed in other QSOs at these redshifts; e.g. \citealt{Mazzucchelli2017,Belladitta2022}) this size corresponds to $\sim$400 Schwarzschild radii ($R_{\rm Sc}$)\footnote{Defined as $R_{\rm Sc}$={2$GM$} / {c$^2$}, where G is the gravitational constant, $M$ is the object mass, and c is the speed of light.}. This limit is consistent with the typical transverse size observed in low redshift jets up to a few thousand $R_{\rm Sc}$ from the central SMBH \citep[e.g.][]{Hada2013}.
Detailed studies in the local Universe have shown that at these scales relativistic jets are still accelerating  and undergoing collimation \citep[e.g.][]{Mertens2016}. At the same time, they can also show signs of mild relativistic boosting \citep[e.g.][]{Kim2018b}. We can therefore use the Compton catastrophe limit on the brightness temperature ($T_\mathrm{B}$ \textless \, 10$^{12}$~K; \citealt{Kellermann1969}) to determine whether relativistic beaming plays an important role in the radio emission and variability of \VIK. Following \cite{Miller-Jones2008}, the brightness temperature, and thus the Compton catastrophe limit, can be written as follows:
\begin{equation}
         T_\mathrm{B} = \frac{\Delta S_\nu  \, (1+z) \,  c^2}{2k \, \nu^2 \, \Omega} \: \leq 10^{12} K,
\label{eq:compt_limit}
\end{equation}
where $\Delta$S$_\nu$ is the amplitude of the variation (0.58~mJy), $\nu$ is the observed frequency (888~MHz), and $c$ and $k$ are the speed of light and Boltzmann's constant, respectively. $\Omega$ is the solid angle subtended by the source and can be expressed as $\Omega = r^2 / d_{\rm A}^2$, where $r$ is the source linear size and $d_{\rm A}$ the angular distance to the source (1131~Mpc). 
Since \VIKs is not resolved in any radio image, we assume the upper limit given by the light-crossing distance ($r$ = 0.04~pc). With these values we find a brightness temperature of $T_{\rm B} \approx 1.4 \times 10^{14}~K$, which is a lower limit to the actual value since we do not know the exact dimensions of the emitting region. From this estimate, we can now derive the corresponding limit on the relativistic Doppler factor \footnote{$\delta = 1 / \Gamma \, (1 - \beta \,\mathrm{cos}\theta_\mathrm{v}$), where $\Gamma$ is the bulk Lorentz factor of the emitting region, $\beta$=$v/c$ and $\theta_\mathrm{v}$ is the viewing angle.} ($\delta$), where the observed brightness temperature value scales as $\delta^3 T_\mathrm{intrinsic}$ \citep{Lahteem1999}. For \VIKs we find that $\delta \gtrsim$5, which would constrain the Doppler factor to values normally observed in radio QSOs whose jets are closely aligned to our line of sight, such as blazars (e.g. \citealt{Hovatta2009,Zhang2020,Homan2021}). This indicates that relativistic beaming might play an important role in the observed radio emission of \VIK.
Moreover, in this case we would also expect the source to have a flat radio spectrum ($\alpha_\mathrm{r}\lesssim$0.5; e.g. \citealt{Padovani2017,Caccianiga2019}), which is in contrast with the slope measured for \VIK: $\alpha_\mathrm{r}$=1.24. Nevertheless, a few exceptions are known where the high-redshift source, despite showing signs of relativistic beaming (with bulk Lorentz factors smaller that the average), presents a relatively steep spectrum (e.g. \citealt{An2020, Spingola2020}).

Another possibility, independent of orientation effects, is that the observed variation could be due to the evolution of the jet itself and its expansion in the circumnuclear medium. If \VIKs is indeed a very young radio source (as hinted by the potential presence of a turnover in the radio spectrum; see section \ref{sec:radio_spec}), we expect its emission to vary as the jets grow in size (e.g. \citealt{Nyland2020}). In particular, we expect the flux density to change mainly close to the peak of the spectrum, as observed in \VIK, with an increase in the optically thick part (i.e. lower frequencies than the turnover) and a decrease in the optically thin part (e.g. \citealt{Orienti2016}). Nevertheless, in order to observe variations of a factor $\sim$2, the typical timescale would be at least several months ($P_\mathrm{radio}\propto t^{2/5}$, e.g. \citealt{An2012}), which is larger than the range sampled by our data. Therefore, it is likely that variations related to the expansion of a young jet, if present, are only of secondary importance (unless strong relativistic beaming is present, which would increase the rest-frame time by a factor $\delta$).

\subsection{Extrinsic Variability}
Another possibility is that the observed variability in \VIKs is caused by refractive interstellar scintillation (RISS). This process occurs when radio waves from sources propagate through the interstellar medium (ISM) of our Galaxy (see e.g. \citealt{Narayan1999} for a detailed description). Since the ISM is composed of charged plasma, variation in its density will deform the wave-front of the incoming radiation from a compact background object, causing a phase delay that can lead to a variation in the observed flux density. The amount and the timescale of the variation depend on the column density of the electrons within the Milky Way along the line of sight (i.e. the position in the sky, $l$=15.942$^{\rm o}$ and $b$=$-$69.303$^{\rm o}$ in Galactic coordinates) as well as the observing frequency.

In order to determine the distribution of the electrons in our Galaxy, and therefore the expected amount of variation from RISS, we used the NE2001 code developed by \cite{Cordes2002,Cordes2003}\footnote{We made use of the python version of the code `pyne2001', available at: https://pypi.org/project/pyne2001/}. In particular, with this code we derived the transitional frequency ($\nu_0$) that separates the strong scattering regime (below $\nu_0$) from the weak one. At the position of \VIKs we obtain $\nu_0\sim$7.9~GHz, which means that at the observed frequency of 888~MHz we are sensitive to strong RISS. Based on the equation in Sec. 3.2.1 in \cite{Walker1998}, the variations corresponding to this transitional frequency have a modulation index of $\sim$29\% and typical timescales of $\sim$10~days. Even though the separation between our data is larger (one year between each data point; two years in total), this flux density variation is consistent with the modulation index measured (both debiased and not) in \VIK. It is therefore likely that the observed variation of \VIKs is mainly related to the scintillation of the ISM in our own Galaxy rather than an intrinsically variable radio jet. However, given the few data available, a systematic monitoring of \VIKs at different frequencies will be essential to better understand the variable nature of this source.

If RISS is the main mechanism responsible for the flux density variation in \VIK, we can set a qualitative limit on the size of the region potentially related to RISS ($\theta_\mathrm{\rm s}$). Indeed for refractive scintillation, above a certain angular size cutoff $\theta_\mathrm{lim}$ (which depends on the position on the sky), angular broadening will take place and the scintillation as well as the modulation index will be quenched by a factor ($\theta_\mathrm{lim}$/$\theta_\mathrm{\rm s}$)$^{7/6}$.
At 888~MHz and at the position of \VIKs, $\theta_{\rm lim}\sim$1~mas (which corresponds to a projected size of $\sim$5~pc).
This means that even if RISS is responsible for variability, the radio emitting region in \VIKs should be concentrated in an area of similar spatial extent, which is smaller than typical sizes of newly born radio galaxies \citep[e.g.][]{Polatidis2003}. Therefore, also in this case, projection effects related to a jet closely oriented to the line of sight must take place.


\section{Discussions and Conclusions}
Having a more reliable knowledge of the radio spectrum of \VIKs with respect to \cite{Ighina2021}, we can now better constrain its radio luminosity and radio-loudness parameter in order to compare it to other QSOs. Based on the curved power law obtained from the fit of all the available data, the average radio luminosity at a rest-frame frequency of 5~GHz is $L_{\mathrm{5GHz}}$=5.5$\pm 1.3 \, \times$10$^{25}$~W~Hz$^{-1}$, which corresponds to a radio loudness of $R_{\mathrm{4400\AA}}$=31.2$\pm {7.5}$ or $R_{\mathrm{2500\AA}}$=42.3$\pm 10.1$, where the errors were computed using the debiased modulation index at 888 MHz. According to these values, \VIKs is among the faintest RL QSOs at high redshift in the radio band (see e.g. Tab. 6 in \citealt{Banados2021}).

Although subject to variability, its faintness could also be related to the young age of the radio jet. Indeed the steep radio spectrum of the source and the presence of a turnover at a few GHz in the rest frame are typical features of compact radio jets in their first expanding phases \citep{Odea1998,Odea2021}. 
The linear size derived for \VIKs from the \cite{Orienti2014} relation for PS radio sources ($\sim$30~pc) is larger than the limits we found through variability arguments in the previous section. This could mean that the majority of the emission is concentrated in the innermost regions of the source, even if the jets extend up to tens of parsecs, as observed in other radio jets \citep[e.g.][]{Turner2012}. Moreover, the orientation of the jet might play an important role on the observed size and therefore the derived values are only indicative. From the highest-resolution image we can only set an upper limit on the source's size of about 8.3~kpc (from the major axis of the ATCA follow-up at 5.5~GHz and eq. 14.5 in \citealt{Fomalont1999}). Hence, Very Long Baseline Interferometry (VLBI) is needed to firmly constrain the dimensions of the source.

Moreover, we further investigated the variability around 888~MHz in the observed frame, which has a significance of 2.5--3$\sigma$. If confirmed, the corresponding varying region is likely very compact in spatial extent (smaller than $\sim$5 parsec),
regardless of its origin. The amplitude of the flux density variation observed in \VIKs is consistent with the expectations from RISS; however, we cannot exclude that it may be also related to an intrinsic change of the intensity of the source. In this case, the corresponding size of the varying region could even be of the order of$\sim$400~$R_{\rm Sc}$. Nevertheless, we stress that current data are not enough to conclusively determine the origin of such variability; further radio observations at different frequencies and timescales along with complementary X-ray data would be required.


\cite{Banados2021} recently discovered a RL QSO at $z$=6.82, PSO~J172.3556+18.7734, which has similar radio characteristics (slope and luminosity; also see \citealt{Momjian2021}) to \VIK. Interestingly, this source also varied by a factor $\sim$2 at 1.4~GHz, even though in this case the flux diminished on a considerably longer timescale (17 years in the observed frame). At the same time, the non-detection of this source in the TGSS and RACS surveys implies that its radio spectrum flattens below 1~GHz in the observed frame, which could also be associated with the presence of a turnover. Together with a kinematic age similar to \VIKs ($\sim$1700~years; \citealt{Momjian2021}), its properties suggest that also in this case the radio emission originates from a relatively young jet. Similar conclusions were also reached for the majority of the other $z$\textgreater5 RL QSOs (e.g. \citealt{Frey2008,Frey2011,Shao2021}).

The existence of \VIKs and the other few known PS sources at \textit{z}$\gtrsim$6 with similar radio properties (discovered from optical/near-IR surveys) suggests that they are part of a much larger population of newly born radio jets present at the end of (and possibly during) the epoch of re-ionisation. In addition, the curvature of the radio spectrum of \VIKs is consistent with the criteria adopted by \cite{Drouart2020} in order to efficiently identify high-redshift radio sources, even though their selection method focuses on lower frequencies (70--230~MHz, within the observable range of the MWA) and would thus miss this source. However, a similar approach could be used to select high-frequency PS objects by combining multi-frequency data points from ongoing and upcoming radio surveys, such as the low, mid and high bands of RACS (700--1800~MHz; \citealt{McConnell2020}) as well as higher frequency surveys (VLASS 3~GHz; \citealt{Lacy2020}) surveys.


\begin{acknowledgements}
We want to thank the anonymous referee for their corrections and suggestions that improved the quality of the paper.\\
In this work we have used data from the ASKAP observatory. The Australian SKA Pathfinder is part of the Australia Telescope National Facility which is managed by CSIRO. Operation of ASKAP is funded by the Australian Government with support from the National Collaborative Research Infrastructure Strategy. ASKAP uses the resources of the Pawsey Supercomputing Centre. Establishment of ASKAP, the Murchison Radio-astronomy Observatory and the Pawsey Supercomputing Centre are initiatives of the Australian Government, with support from the Government of Western Australia and the Science and Industry Endowment Fund. We acknowledge the Wajarri Yamatji people as the traditional owners of the Observatory site.\\
This scientific work makes use of the Murchison Radio-astronomy Observatory, operated by CSIRO. We acknowledge again the Wajarri Yamatji people as the traditional owners of the Observatory site. Support for the operation of the MWA is provided by the Australian Government (NCRIS), under a contract to Curtin University administered by Astronomy Australia Limited. We acknowledge the Pawsey Supercomputing Centre which is supported by the Western Australian and Australian Governments.\\
This paper includes archived data obtained through the CSIRO ASKAP Science Data Archive, CASDA (\url{http://data.csiro.au}).\\ 
The Australia Telescope Compact Array is part of the Australia Telescope National Facility which is funded by the Australian Government for operation as a National Facility managed by CSIRO. We acknowledge the Gomeroi people as the traditional owners of the Observatory site.\\
This paper makes use of the following ALMA data: ADS/JAO.ALMA\#2019.1.00147.S. ALMA is a partnership of ESO (representing its member states), NSF (USA) and NINS (Japan), together with NRC (Canada), MOST and ASIAA (Taiwan), and KASI (Republic of Korea), in cooperation with the Republic of Chile. The Joint ALMA Observatory is operated by ESO, AUI/NRAO and NAOJ.\\
LI, SB, AC, AM acknowledge financial contribution from the agreement ASI-INAF n. I/037/12/0 and n.2017-14-H.0 and from INAF under PRIN SKA/CTA FORECaST.\\
JL and JP are supported by Australian Government Research Training Program Scholarships. 
YW is supported by the China Scholarship Council.\\
This research made use of the Common Astronomy Software Applications package (CASA), \cite{Mcmullin2007}.
This research made use of Astropy (\url{http://www.astropy.org}) a community-developed core Python package for Astronomy \citep{astropy2018}.

\end{acknowledgements}

\bibliographystyle{aa} 
\bibliography{biblio} 

\appendix

\section{Radio images}
\label{sec:radio_images}
\begin{figure*}
    \centering
    \includegraphics[width=0.33\linewidth]{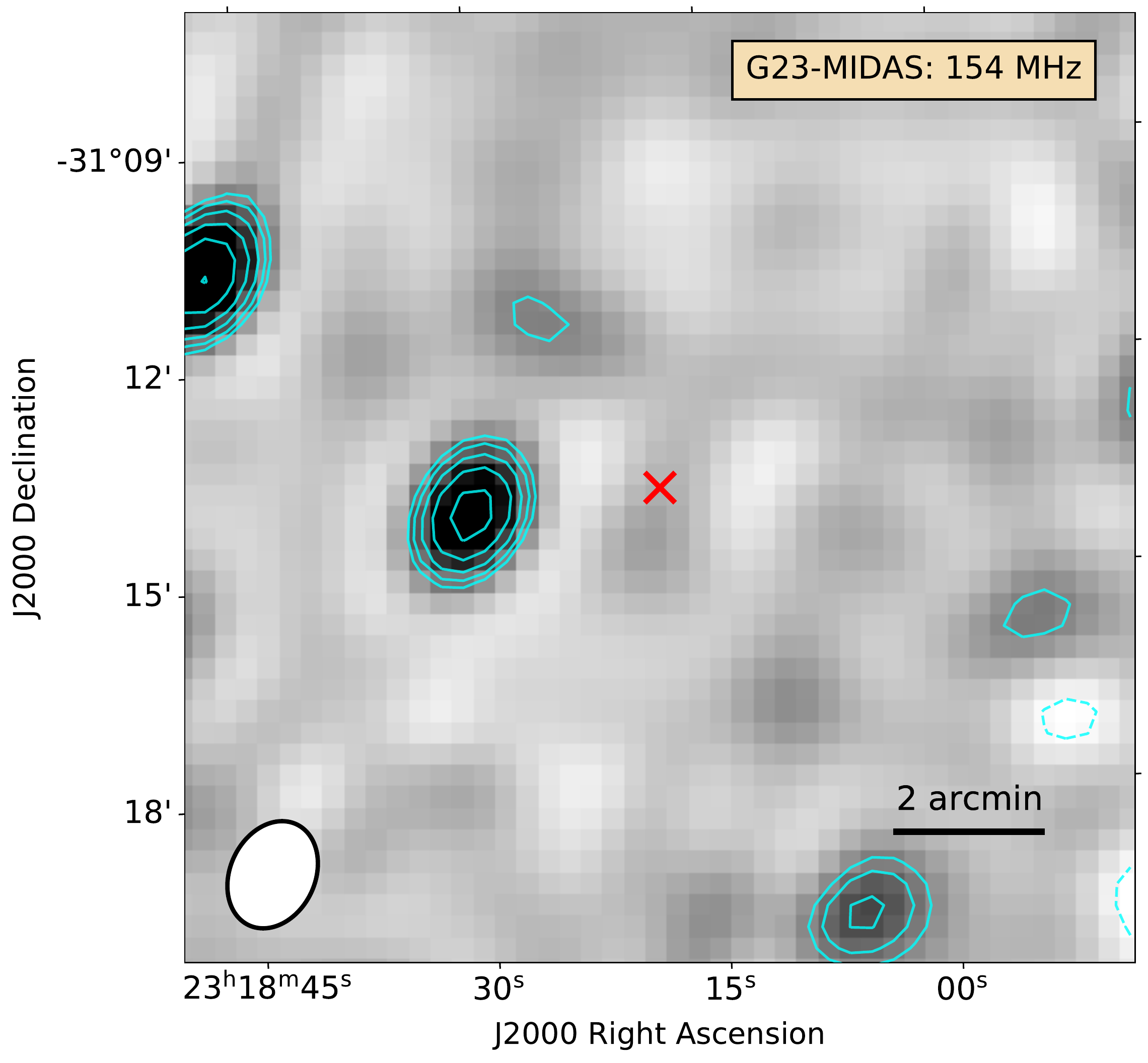}
    \includegraphics[width=0.33\linewidth]{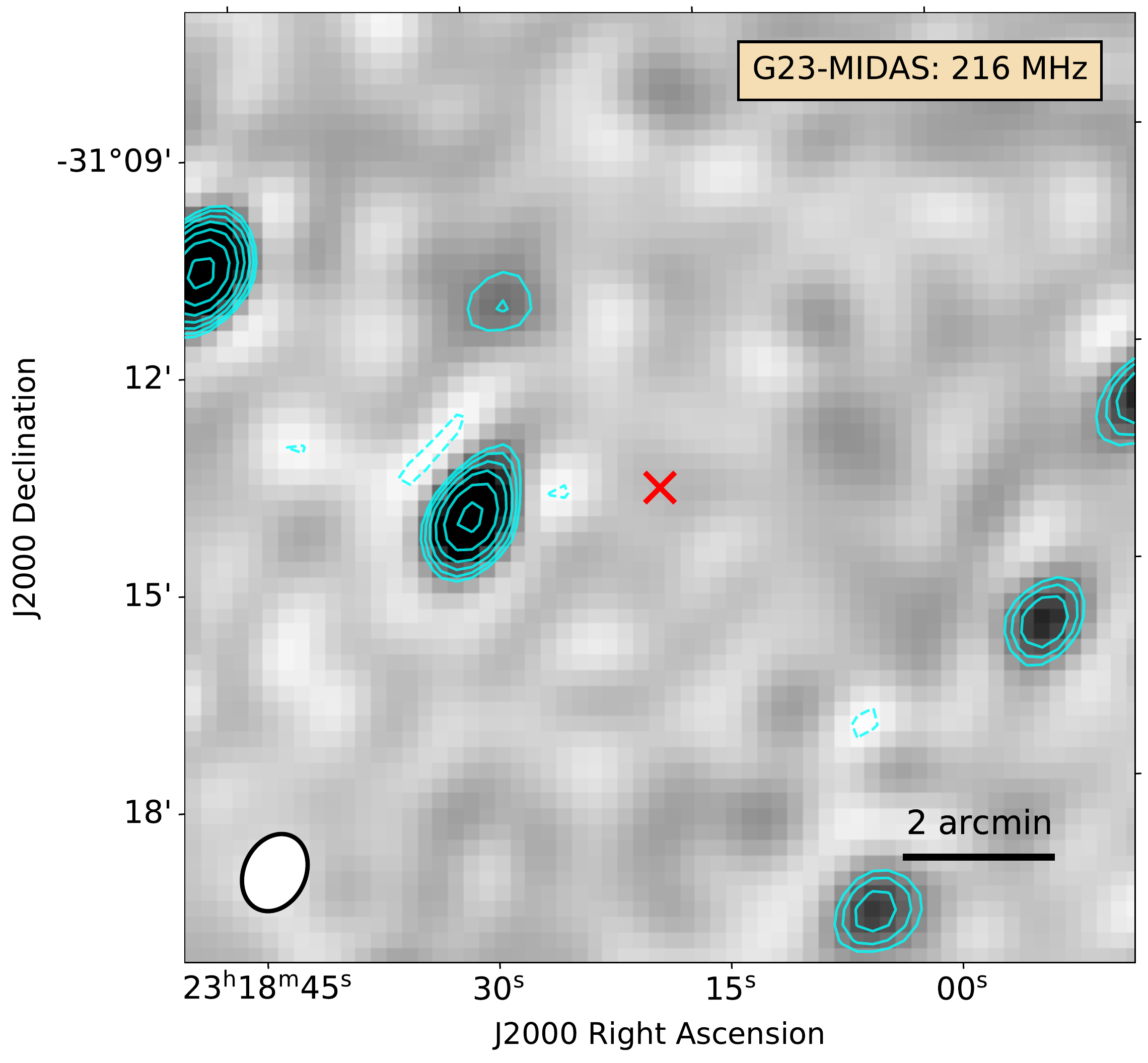}\\      \includegraphics[width=0.33\linewidth]{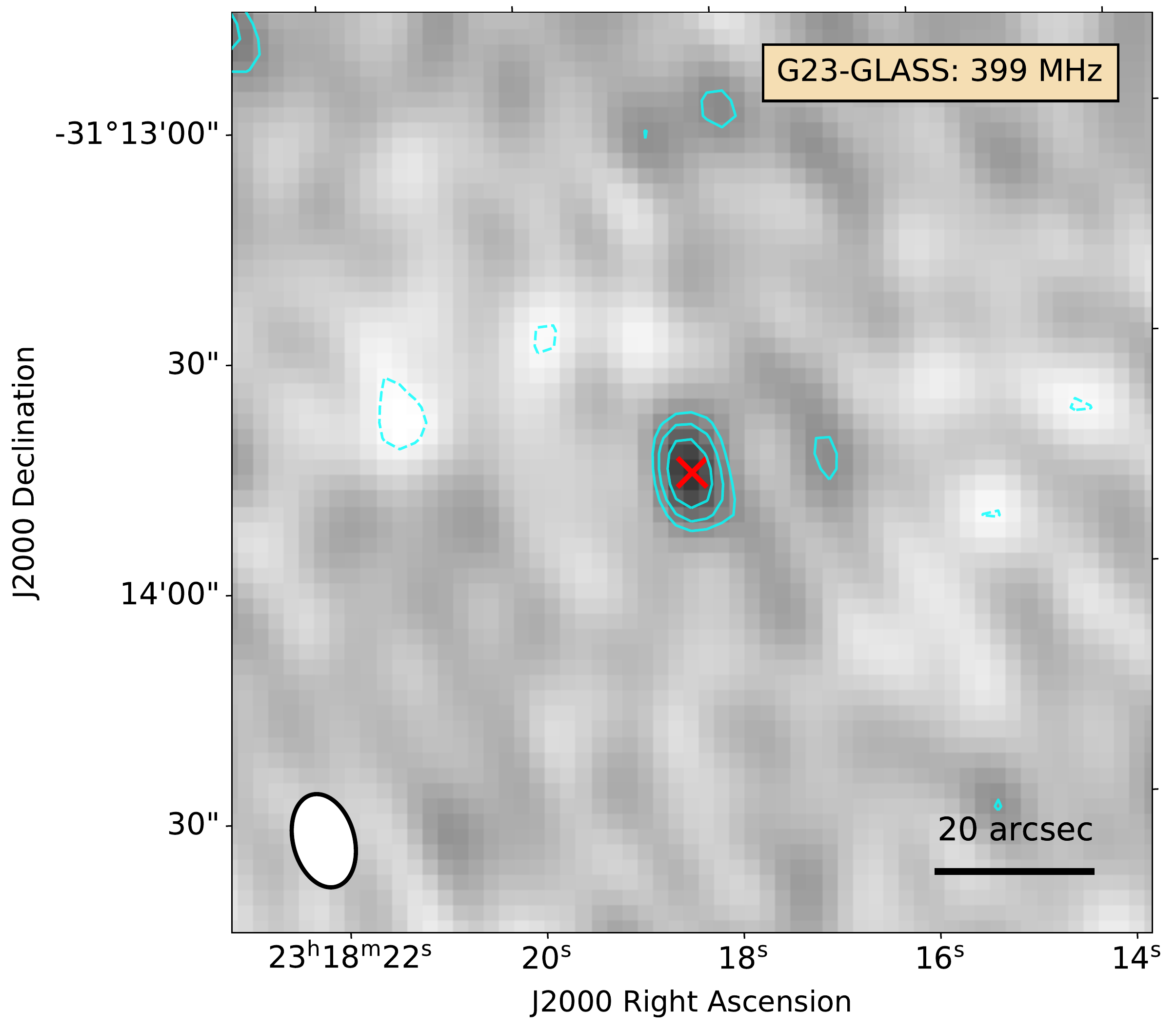}
    \includegraphics[width=0.33\linewidth]{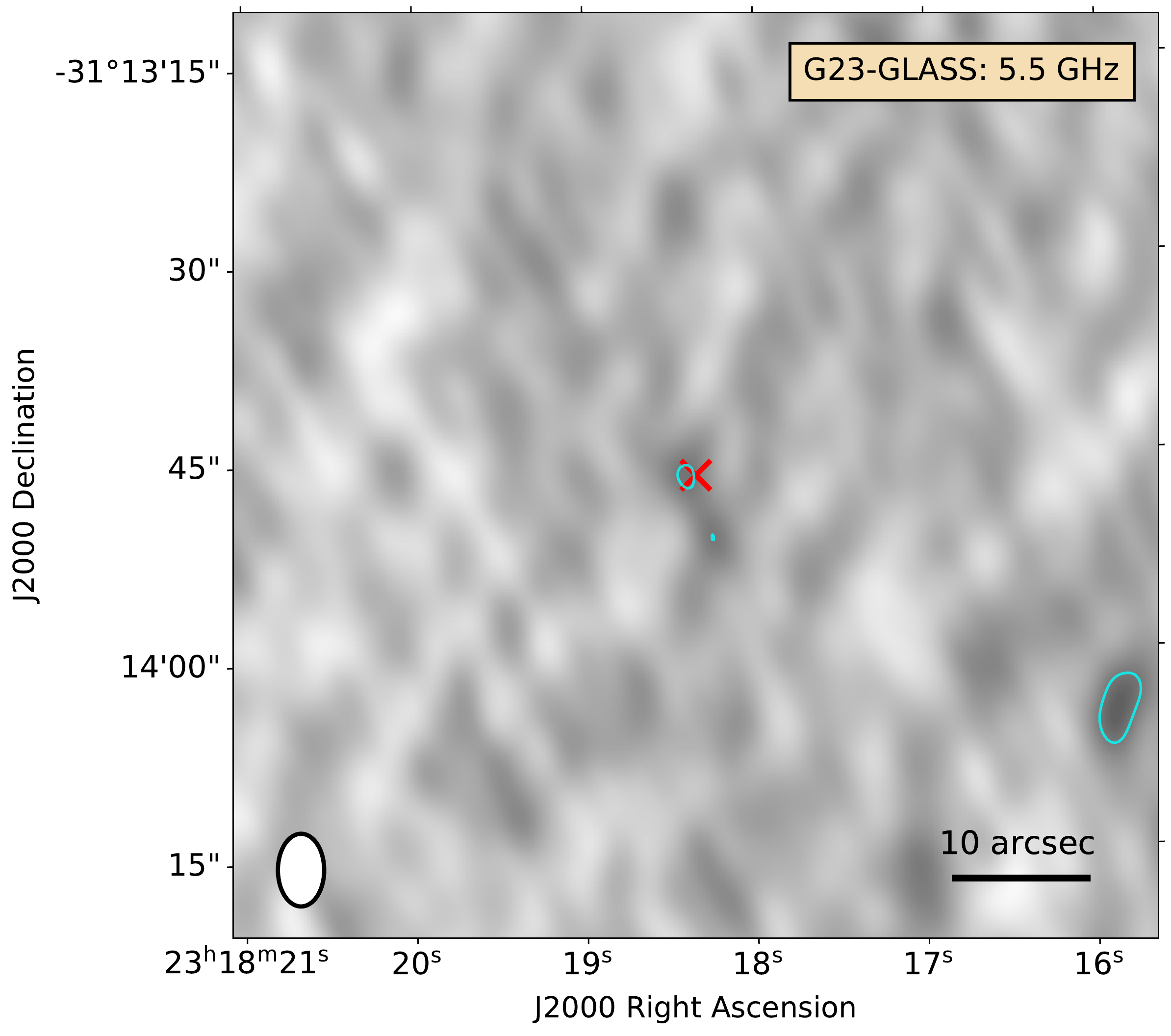}
    
\caption{Radio images of \VIKs from the G23 field. In all the images, the radio contours shown in cyan are spaced by $\sqrt{2}$ starting from $\pm$3 $\times$ RMS (solid and dashed lines indicate positive and negative contours respectively). The position of the optical counterpart is marked with a red cross and the synthesised beam of each observation is shown in the bottom left-hand corner.} 

\label{fig:G23_images}
\end{figure*}

\begin{figure*}
    \centering
    \includegraphics[width=0.33\linewidth]{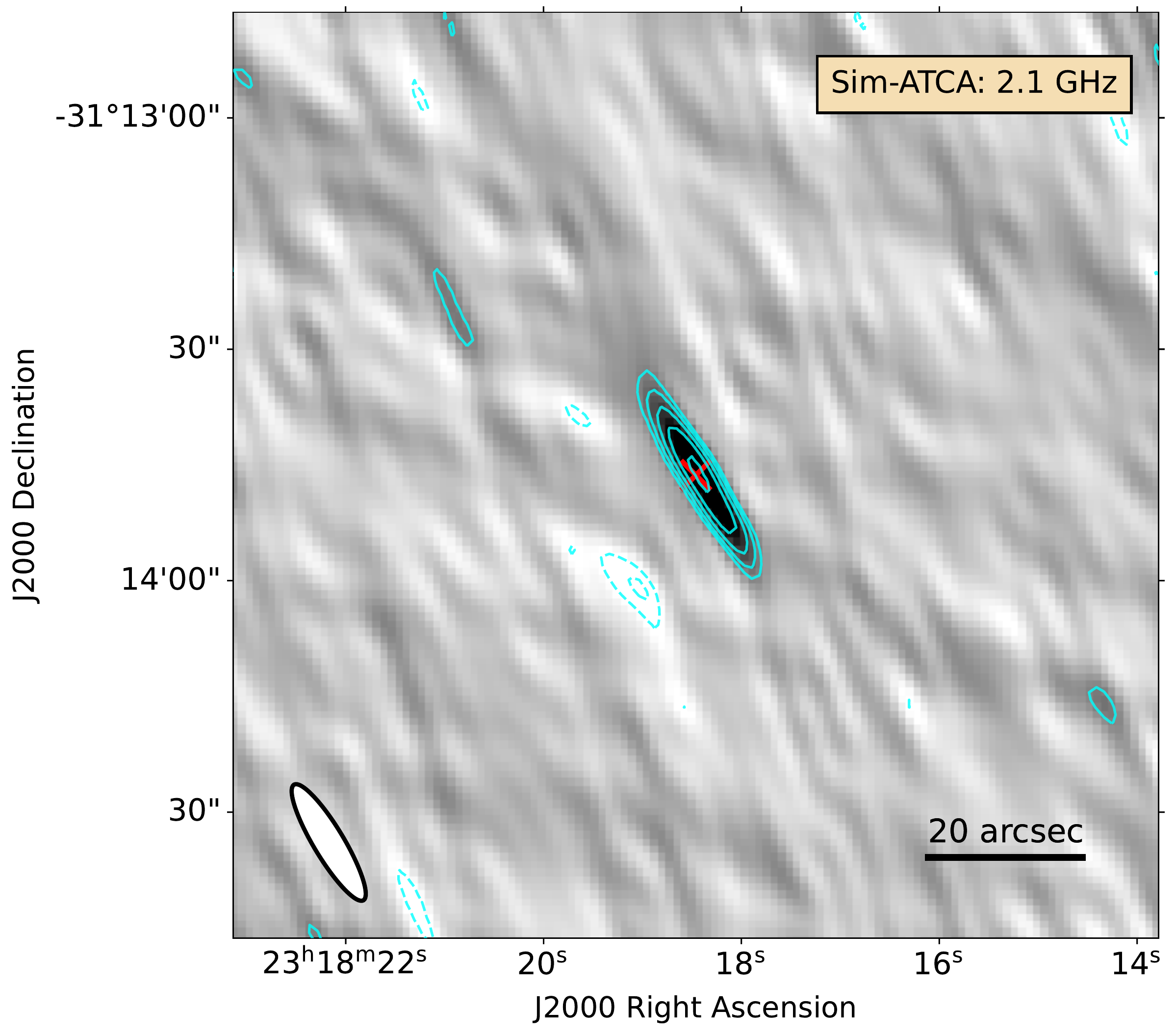}    
    \includegraphics[width=0.33\linewidth]{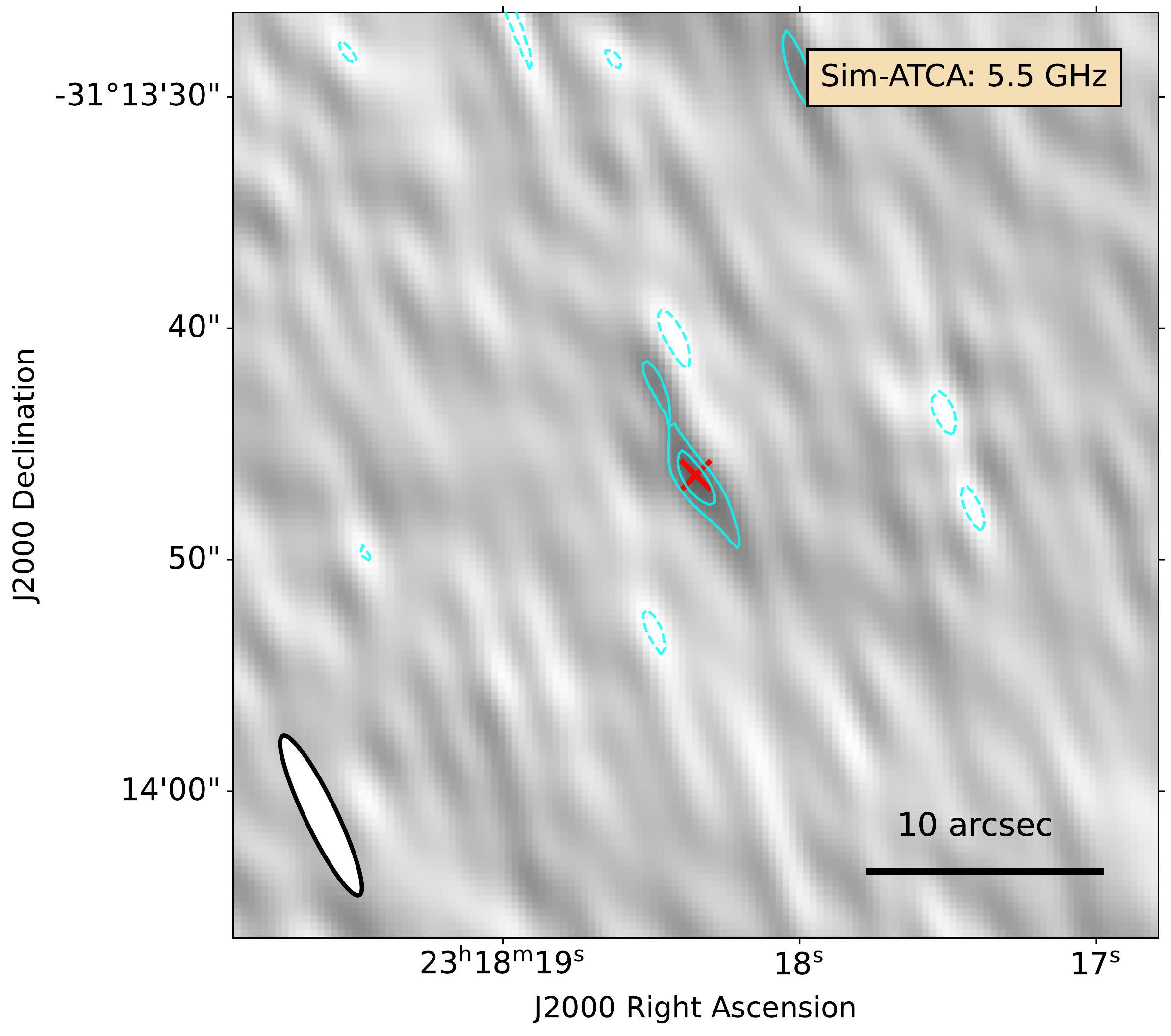}
    \includegraphics[width=0.33\linewidth]{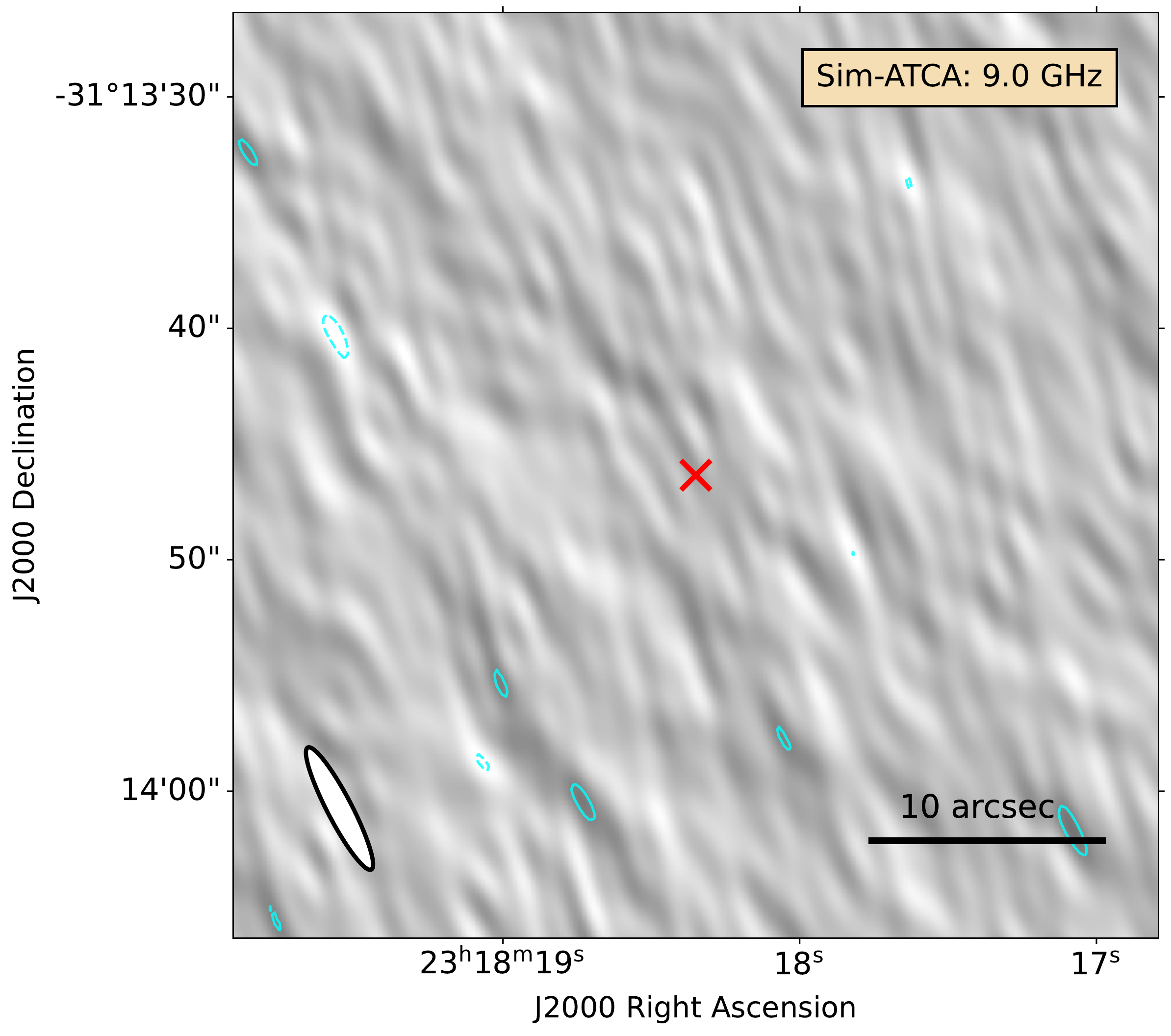}
    \includegraphics[width=0.33\linewidth]{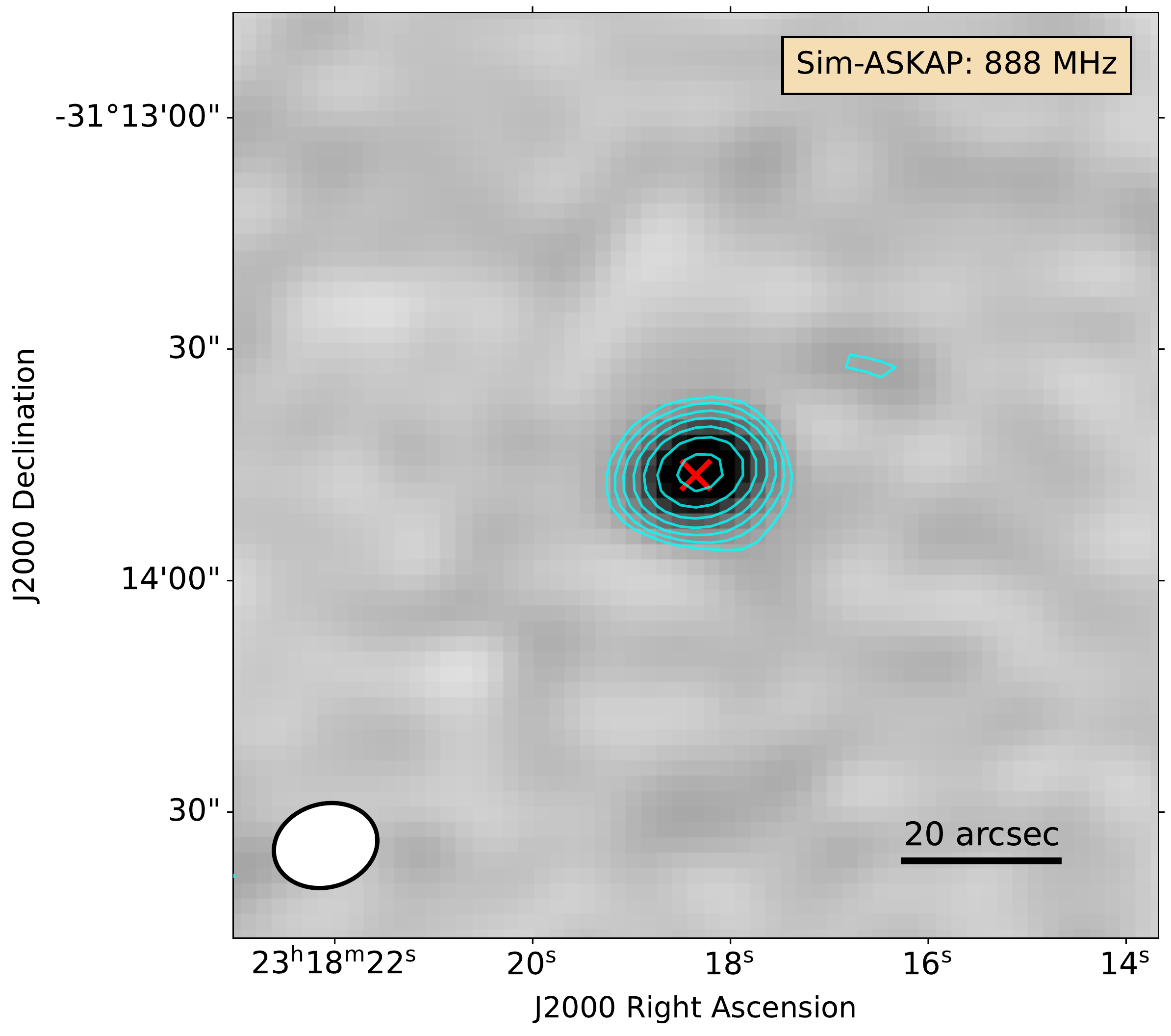}
    \includegraphics[width=0.33\linewidth]{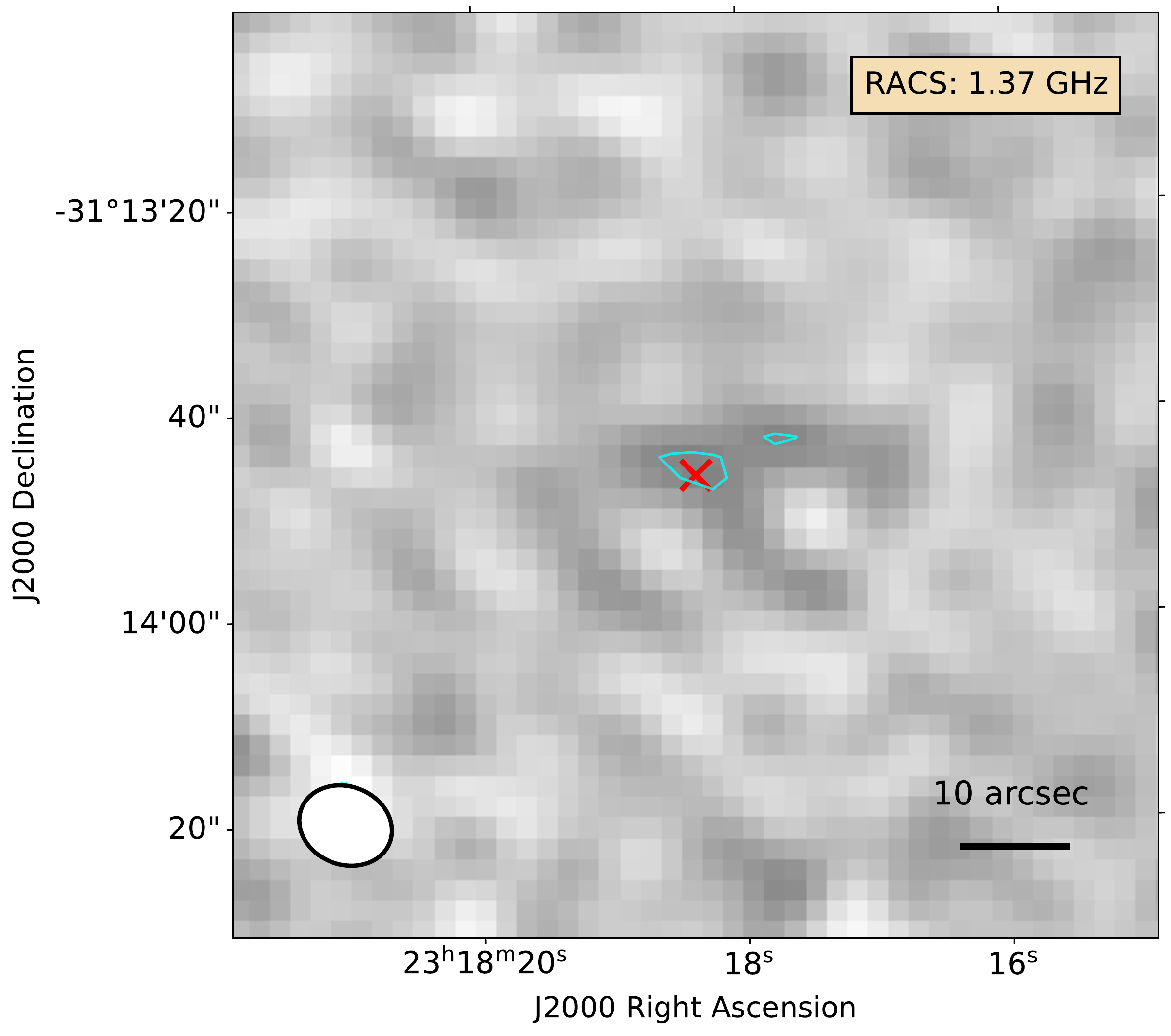}

\caption{Same as Fig. \ref{fig:G23_images}, but for the simultaneous follow-up radio observations and the mid-band (1.37~GHz) scan of the RACS survey.}
         
\label{fig:too_images}
\end{figure*}

In this section we report the radio images of \VIKs from the G23 radio surveys (Fig. \ref{fig:G23_images}) and the follow-up and RACS-mid observations (Fig. \ref{fig:too_images}).

\end{document}